\documentclass[prc,twocolumn,showpacs,floatfix,superscriptaddress]{revtex4}
\usepackage{hyperref}
\usepackage{graphicx}
\usepackage{amssymb}
\usepackage{amsmath}
\usepackage{txfonts}
\usepackage{enumerate}

\newcommand{\calcium}[1][40]{{}^{#1}\mathrm{Ca}}

\begin{document}

\title{Relevance of equilibrium in multifragmentation}

\author{Takuya Furuta}
\altaffiliation[Present address:]{
LPC Caen (CNRS-IN2P3/ENSICAEN et Universit\'e), F-14050 Caen, France.
}                                                                   
\affiliation{
GANIL (DSM-CEA/IN2P3-CNRS), B.P.55027, F-14076 Caen, France
}
\author{Akira Ono}
\affiliation{
Department of Physics, Tohoku University, Sendai
980-8578, Japan
}

\begin{abstract}
  The relevance of equilibrium in a multifragmentation reaction of
  very central $\calcium+\calcium$ collisions at 35 MeV/nucleon is
  investigated by using simulations of antisymmetrized molecular
  dynamics (AMD). Two types of ensembles are compared. One is the
  reaction ensemble of the states at each reaction time $t$ in
  collision events simulated by AMD, and the other is the equilibrium
  ensemble prepared by solving the AMD equation of motion for a
  many-nucleon system confined in a container for a long time. The
  comparison of the ensembles is performed for the fragment charge
  distribution and the excitation energies. Our calculations show that
  there exists an equilibrium ensemble that well reproduces the
  reaction ensemble at each reaction time $t$ for the investigated
  period $80\leq t\leq300$ fm/$c$. However, there are some other
  observables that show discrepancies between the reaction and
  equilibrium ensembles. These may be interpreted as dynamical effects
  in the reaction. The usual static equilibrium at each instant is not
  realized since any equilibrium ensemble with the same volume as that
  of the reaction system cannot reproduce the fragment observables.
\end{abstract}

\maketitle
\bibliographystyle{physics}

\section{Introduction}
\label{INTRODUCTION}
In medium-energy heavy-ion collisions at around the Fermi energy,
intermediate-mass fragments as well as a large number of light
particles such as nucleons and alpha particles are copiously produced
\cite{montoya,lukasik,lefort,reisdorf,pochodzalla}. This phenomenon is
called multifragmentation. It is a challenging problem to understand
the complex but rich quantum many-body dynamics of multifragmentation.
One of the purposes of studying heavy-ion collisions is to explore the
properties of nuclear matter \cite{danielewiczSci,fuchsWCI}. This
information is valuable not only for nuclear physics but also for
astrophysical interests such as supernova explosions and the structure
of neutron stars \cite{horowitzWCI}. The nuclear matter is expected to
be compressed in the initial stage of a collision and the created
compressed matter then expands afterward. The study of heavy-ion
collisions thus offers a possibility to probe the properties of
nuclear matter in a wide range of density. Multifragmentation has been
considered to occur in the expanding stage and to have some connection
to the nuclear liquid-gas phase transition, the existence of which is
speculated based on the resemblance between the equation of state of
homogeneous nuclear matter and that of a van der Waals system.
Intensive research has been carried out to find evidence of this phase
transition in experimental data of multifragmentation.  In some works
it is claimed that indications have been obtained
\cite{guptaExpEvi,pochodzalla,chomaz,agostinoNP,agostinoFLCT,natowitz,elliott,trautmann}.
However, they are not conclusive and much effort is still required.

One of the difficulties is that it is not straightforward to relate
the experimental data of heavy-ion collisions with the statistical
properties of nuclear matter unless the state variables such as the
temperature are well defined in dynamical reactions. The typical
reaction time scale of multifragmentation reactions is the order of
100 fm/$c$, which may not be long enough for the system to reach
equilibrium compared with the typical time scale of successive
two-nucleon collisions (a few tens of fm/$c$). However, there are
several reports that support the achievement of equilibrium. An
example is the existence of several types of scaling laws that appear
in experimental data (e.g.\ Fisher's scaling \cite{elliott} and
isoscaling \cite{Xu,tsangIsoscaling}), which may be understood if the
system has reached equilibrium. Another example is the reasonable
reproduction of the fragment mass (charge) distribution by statistical
models for some multifragmentation reactions
\cite{randrupSM,bondorfSM,agostinoSM,grossPR,botvinaWCI}. However, the
achievement of equilibrium in multifragmentation reactions is still a
controversial issue. One of the difficulties is that the information
obtained directly from experiments is that of the very last stage of
the reactions. Even if the system reaches equilibrium, the system
undergoes the sequential decay process that distorts the information
at the stage of the equilibrium before the fragments are finally
detected in experiments. Another difficulty is that even if the
equilibrium is relevant to multifragmentation reactions the
achievement can be incomplete. Several aspects are expected to reflect
the reaction dynamics, such as the pre-equilibrium emissions of light
particles, the collective flow, and the expansion of the system
\cite{marie,desesquelles,reisdorfAnnu,wada1998,andronicWCI}.

The aim of this paper is to investigate whether the concept of
equilibrium is relevant in multifragmentation, and if so, in what
sense. We examine the achievement of equilibrium in multifragmentation
reactions simulated by antisymmetrized molecular dynamics (AMD)
\cite{onoPRL,onoPTP,onoReview}. AMD is a microscopic dynamical model
based on the degrees of freedom of interacting nucleons. AMD is a
suitable model for this study for the following reasons: It has been
shown that various aspects of experimental data are reproduced by
applying AMD to nuclear reactions
\cite{onoPRL,onoPTP,onoAMD-V,onoAu,onoWithFrance,onoReview,wada1998,wada2000,wada2004,hudan}.
It has been also argued that the quantum and fermionic statistical
properties of nuclear systems are correctly described by AMD if an
appropriate quantum branching process is taken into account
\cite{ohnishiPRL,ohnishiAP,onoAMDMF,onoSts}.  Furthermore, we can
construct microcanonical equilibrium ensembles for given energies and
volumes by solving the AMD equation of motion of a many-nucleon system
confined in a container for a long time \cite{furutaLGpt}. By
extracting temperature and pressure from these equilibrium ensembles
and interpolating these data, we have drawn the constant-pressure
caloric curves to show that negative heat capacity, which is a signal
of the phase transition in finite systems
\cite{grossPR,grossEP,grossPCCP}, appears in the obtained result.

To investigate the relevance of equilibrium in multifragmentation, we
employ the following steps.  We perform the AMD simulation for very
central $\calcium+\calcium$ collisions at 35 MeV/nucleon.  The
reaction ensemble at each reaction time $t$ is constructed by
collecting the many-nucleon states at the time $t$ from different
events. We compare this reaction ensemble with an equilibrium ensemble
with appropriately chosen energy and volume. If we are able to find an
equilibrium ensemble that is equivalent to the reaction ensemble, we
may be able to discuss the connection between the multifragmentation
data and statistical properties of nuclear matter.  This subject has
been studied by Raduta \textit{et al.}
\cite{radutaEQ,radutaEQ2}. They have compared an ensemble obtained by
the stochastic mean-field approach \cite{colonna} which is a BUU-type
transport model with that obtained by the microcanonical
multifragmentation model \cite{radutaMMM} which is a statistical
model.  In contrast, we use the same version of AMD to describe both
dynamical and equilibrium situations so that we can compare the
reaction and equilibrium ensembles directly without being affected by
the model difference.

This paper is organized as follows.  In Sec.\ \ref{AMDformalism}, the
framework of AMD, which is used to simulate reaction and equilibrium
systems, is explained. In Sec.\ \ref{REACTION}, we show the results of
the AMD simulation for the $\calcium+\calcium$ collisions at 35
MeV/nucleon, which have been already studied with another version of
AMD \cite{onoAMD-V}. One of the purposes of this section is to ensure
that the modifications introduced in Ref.\ \cite{furutaLGpt} for the
application to statistical calculations do not spoil the good
reproduction of the reaction data. Limiting the discussion to the very
central reaction, we also argue the time evolution of the reaction
system by showing the fragment observables. In Sec.\
\ref{EQUILIBRIUM}, we show results of the statistical calculation for
an equilibrium system with 18 protons and 18 neutrons which is the
same system as Ref.\ \cite{furutaLGpt}. It is confirmed that negative
heat capacity appears in the constant-pressure caloric curves although
several modifications are introduced in this paper.  In Sec.\
\ref{COMPARISON}, we compare the ensembles obtained by the dynamical
simulation (Sec.\ \ref{REACTION}) and obtained by the statistical
calculations with various conditions of volume and energy (Sec.\
\ref{EQUILIBRIUM}), and discuss whether the concept of equilibrium is
relevant to the multifragmentation reaction.  Section \ref{SUMMARY} is
devoted to a summary and future perspectives.

\section{Framework of AMD time evolution}
\label{AMDformalism}
In this section, we present our AMD framework to calculate the time
evolution of many-nucleon systems. We basically follow the framework
of Ref.\ \cite{furutaLGpt}, although several modifications are
introduced in the present work.  We simulate both a multifragmentation
reaction (Sec.\ \ref{REACTION}) and an equilibrium system (Sec.\
\ref{EQUILIBRIUM}) with the same AMD model.

The wave function of an $A$-nucleon system $|\Psi(t)\rangle$ that
evolves with time $t$ according to the many-body Hamiltonian is given
by a superposition of various reaction channels.  As it is impossible
to follow the exact time evolution of $|\Psi(t)\rangle$ in practice,
in the AMD formalism we approximate the many-body density matrix
$|\Psi(t)\rangle\langle\Psi(t)|$ by an ensemble of AMD wave functions
$|\Phi(Z)\rangle$ as
\begin{equation}
|\Psi(t)\rangle\langle\Psi(t)|\approx
\int\frac{|\Phi(Z)\rangle\langle\Phi(Z)|}
{\langle\Phi(Z)|\Phi(Z)\rangle}
w(Z,t)dZ,
\label{eq:AMDensemble}
\end{equation}
where $w(Z,t)$ is the weight factor for each reaction channel at time
$t$.  This approximation implies that we incorporate the existence of
various reaction channels while we ignore the interference between
channels since it is unimportant for practical purposes (decoherence).

AMD uses a single Slater determinant of Gaussian wave packets as a
channel wave function
\begin{equation}
\langle\mathbf{r}_1\cdots\mathbf{r}_A|\Phi(Z)\rangle=\det_{ij}
\Bigl[\varphi_{\mathbf{Z}_i}(\mathbf{r}_j)\chi_{\alpha_i}(j)\Bigr],
\label{eq:AMDWF}
\end{equation}
where the spatial wave functions of nucleons $\varphi_{\mathbf{Z}}$
are given by
\begin{equation}
\langle\mathbf{r}|\varphi_{\mathbf{Z}}\rangle
=\Bigl(\frac{2\nu}{\pi}\Bigr)^{3/4}\exp\Bigl[
-\nu\Bigl(\mathbf{r}-\frac{\mathbf{Z}}{\sqrt{\nu}}\Bigr)^2\Bigr]
\label{eq:Gaussian}
\end{equation}
and $\chi_{\alpha}$ denotes the spin-isospin wave function,
$\chi_{\alpha}=p\uparrow$, $p\downarrow$, $n\uparrow$, and
$n\downarrow$. The AMD wave function $|\Phi(Z)\rangle$ is the
many-nucleon state parametrized by a set of complex variables
$Z\equiv\{\mathbf{Z}_i\}_{i=1,\dots,A}$. The real and the imaginary
parts of $\mathbf{Z}$ correspond to the centroids of the position and
the momentum of each wave packet, respectively, if the
antisymmetrization effect is ignored.  The width parameter $\nu$ is
treated as a constant parameter common to all the wave packets and
$\nu=0.16\;\text{fm}^{-2}$ is utilized in this paper, which has been
adjusted to reasonably describe ground states of light nuclei such as
${}^{16}\mathrm{O}$. It is shown that the binding energies of nuclei
in a wide range of the nuclear chart are reproduced well with
appropriate effective interactions \cite{onoC12,onoReview}. This
choice of channel wave function is suitable for the simulation of
multifragmentation reactions, where each single-particle wave function
should be localized within a fragment. Besides, the AMD wave function
$|\Phi(Z)\rangle$ contains many quantum features owing to
antisymmetrization and so is even utilized for nuclear structure
studies \cite{enyo}.

According to Eq.\ (\ref{eq:AMDensemble}), the time evolution of the
$A$-nucleon system may be determined by calculating the time evolution
of the weight factor for each channel $w(Z,t)$.  Alternatively we take
another viewpoint that the parameters $Z$ of the wave function
$|\Phi(Z)\rangle$ are stochastic time-dependent variables $Z(t)$ and
the time evolution of the many-nucleon state is given by the ensemble
of the various trajectories.  The stochastic time evolution of $Z(t)$
should be considered as the quantum branching from a channel
$|\Phi(Z)\rangle$ to many other channels $|\Phi(Z_1')\rangle$,
$|\Phi(Z_2')\rangle$, \ldots.

The time evolution of the centroids $Z$ is determined by a stochastic
equation of motion symbolically written as
\begin{equation}
\frac{d}{dt}\mathbf{Z}_i=\{\mathbf{Z}_i,\mathcal{H}\} +(\text{NN
collision})+ \Delta\mathbf{Z}_i
\label{eq:EOMsim}.
\end{equation}
The first term $\{\mathbf{Z}_i,\mathcal{H}\}$ is the deterministic
term which is derived from the time-dependent variational principle
\cite{onoPRL,onoPTP,onoReview}. The Gogny force \cite{gogny} is
adopted as the effective interaction and the Coulomb force is also
taken into account. The second term represents the stochastic
two-nucleon collision process where a parametrization of the
energy-dependent in-medium cross section is adopted \cite{onoReview}.

The third term $\Delta\mathbf{Z}_i$ is a stochastic fluctuation term
that has been introduced to compromise the unrestricted
single-particle motion in the mean-field and the localization of
single-particle wave functions at the time of forming fragments
\cite{onoAu,onoWithFrance,onoReview}.  The fluctuation
$\Delta\mathbf{Z}_i$ is determined so that the evolution of the width
and shape of the single-particle phase-space distribution in
mean-field theories is reproduced for a certain time duration
$\tau_\text{cohe}$ by the ensemble average of the localized
single-particle phase-space distribution of each channel.  In
practice, we compute $\Delta\mathbf{Z}_i$ by solving the Vlasov
equation with the same effective interaction used in the term
$\{\mathbf{Z}_i,\mathcal{H}\}$.  The time duration $\tau_\text{cohe}$
to respect the coherent single-particle motion in the mean-field
should be related to many-body effects in some way since the
decoherence is due to the many-body correlations beyond mean-field. In
this paper, we choose $\tau_\text{cohe}$ in such a way that the
decoherence probability becomes approximately proportional to the
density at the nucleon location. This stochastic term is essential for
the consistency of dynamics with quantum statistics
\cite{ohnishiPRL,ohnishiAP,onoAMDMF,onoSts}.

Basically, we follow the formalism explained in Ref.\
\cite{furutaLGpt}.  In the present work, we have chosen the
probability of decoherence for the nucleon $k$ to occur during the
time interval $\Delta t$ to be
\begin{equation}\label{eq:tau0}
P_{\text{dech}\,k}=1-\exp\left(-\frac{\rho_k\Delta
t}{\rho_0\tau_0}\right),
\end{equation}
where $\tau_0$ is chosen to be 5 fm/$c$, $\rho_k$ is the density at
the wave packet center of the nucleon $k$ excluding the contribution
from the nucleon $k$ itself, and $\rho_0=0.16$ $\text{fm}^{-3}$ is the
normal nuclear matter density.  To better describe the
$\calcium+\calcium$ reaction at 35 MeV/nucleon, the dissipation term
(corresponding to the fluctuation term) is assumed to conserve the
monopole and quadrupole moments in coordinate and momentum spaces for
the nucleons that have more than 15 neighboring nucleons as in Ref.\
\cite{onoWithFrance}.  Furthermore, another modification is introduced
to better incorporate the effect of decoherence, the details of which
are given in the Appendix.

\section{Application to a reaction}
\label{REACTION}
We apply AMD to $\calcium + \calcium$ collisions at 35 MeV/nucleon.
This system has been already studied by using AMD with the
instantaneous decoherence \cite{onoAMD-V} and it has been shown that a
good reproduction of experimental data is obtained
\cite{onoAMD-V,wada1998}. However, we take a finite coherence time and
we also introduce some modifications explained in Ref.\
\cite{furutaLGpt} and the Appendix.  Therefore, we confirm the
applicability of the present framework by comparing the simulation
results with the experimental data by Hagel \textit{et al.}
\cite{Hagel}. The ensembles of the many-nucleon states obtained from
the dynamical simulations in this section are utilized in Sec.\
\ref{COMPARISON}.

\begin{figure}
\begin{center}
\includegraphics[width=\columnwidth]{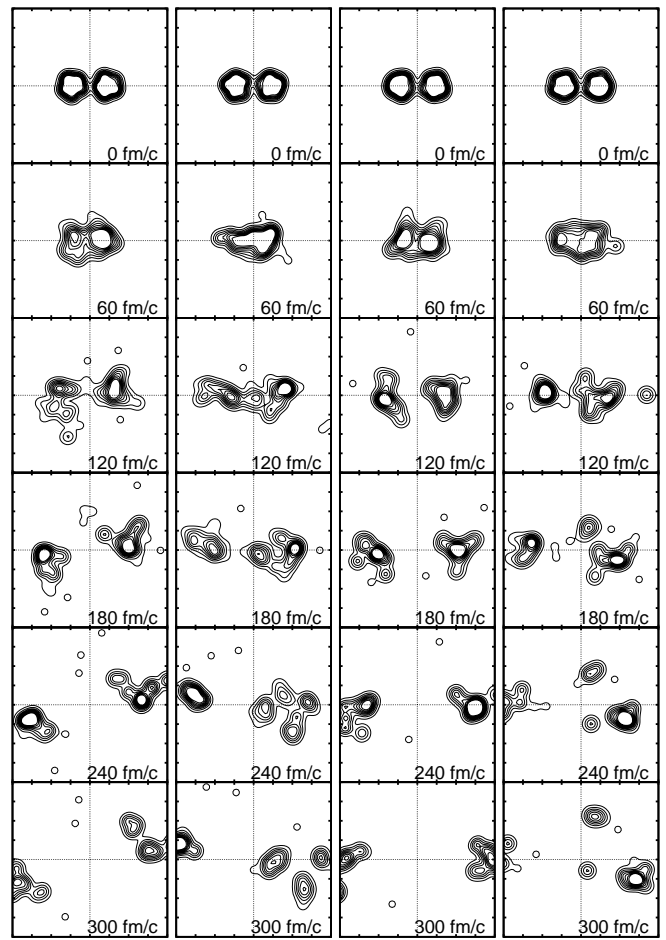}
\end{center}
\caption{Density distributions projected on the reaction plane of very
  central $\calcium+\calcium$ collisions at 35 MeV/nucleon
  ($b_{\text{imp}}=0$ fm) from $t=0$ fm/$c$ to $t=300$ fm/$c$ for four
  different events. The size of the displayed area is $40\times40$
  fm.}
\label{fig:manyeach}
\end{figure}

\begin{figure}
\begin{center}
\includegraphics[width=\columnwidth]{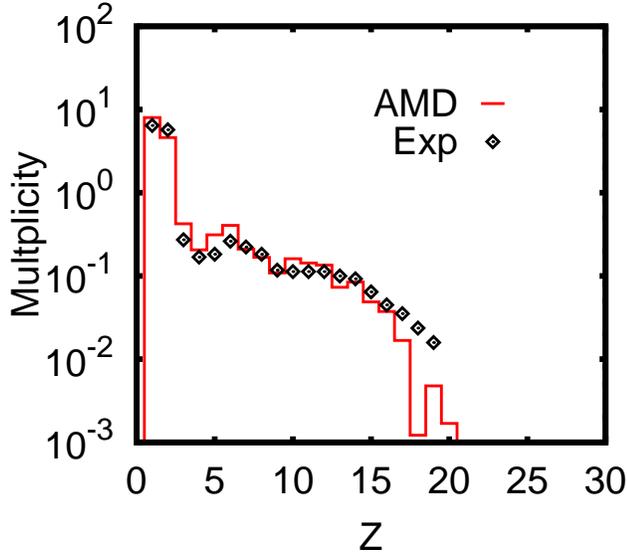}
\end{center}
\caption{(Color online) The fragment charge distribution of the
reaction $\calcium + \calcium$ at 35 MeV/nucleon simulated by AMD
(full line) compared with the
experimental data of Hagel \textit{et al.} \cite{Hagel} (points).}
\label{fig:compareEX}
\end{figure}

\begin{figure}
\begin{minipage}{.47\columnwidth}
\centering
\includegraphics[width=\columnwidth]{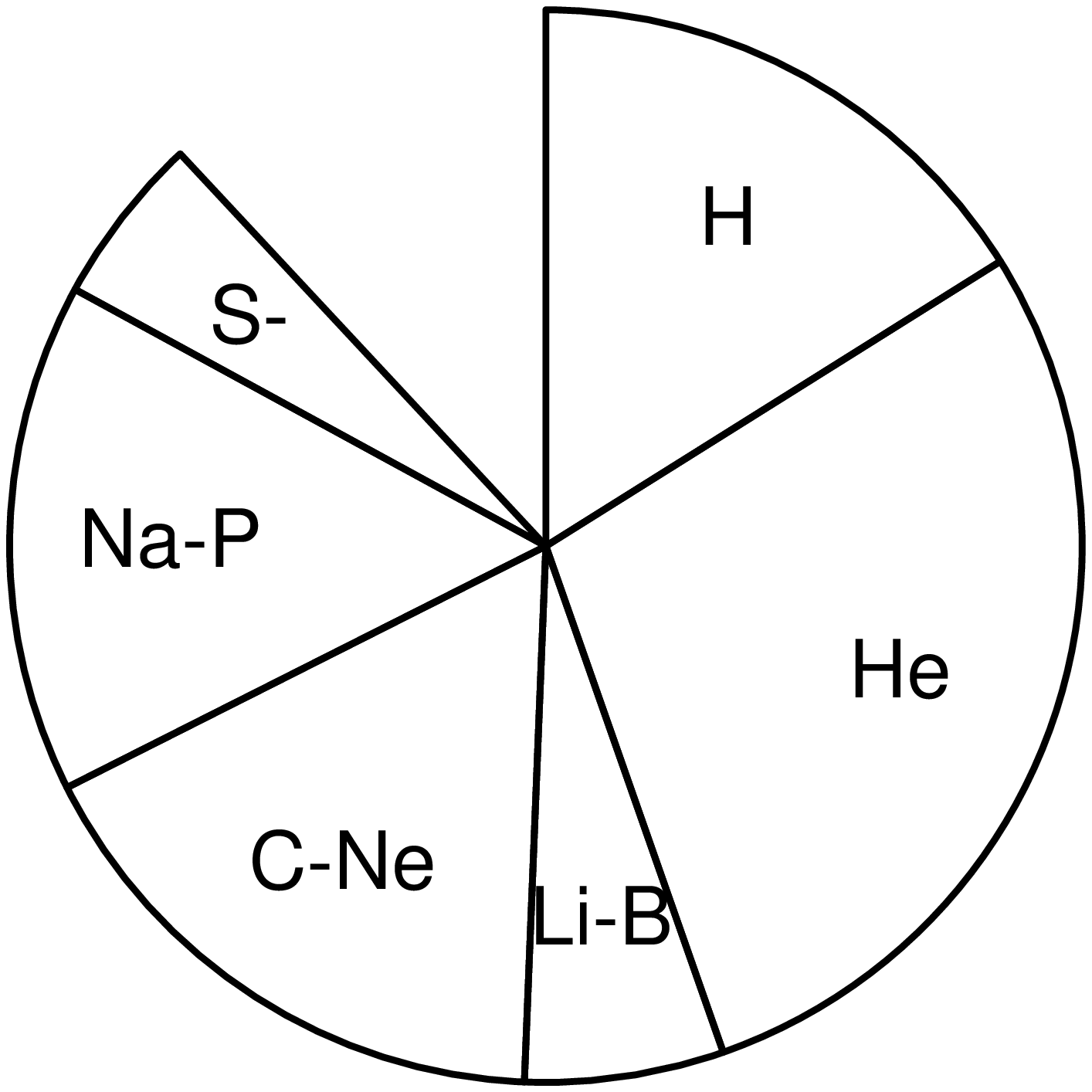}
EXP
\end{minipage}
\hfill
\begin{minipage}{.47\columnwidth}
\centering
\includegraphics[width=\columnwidth]{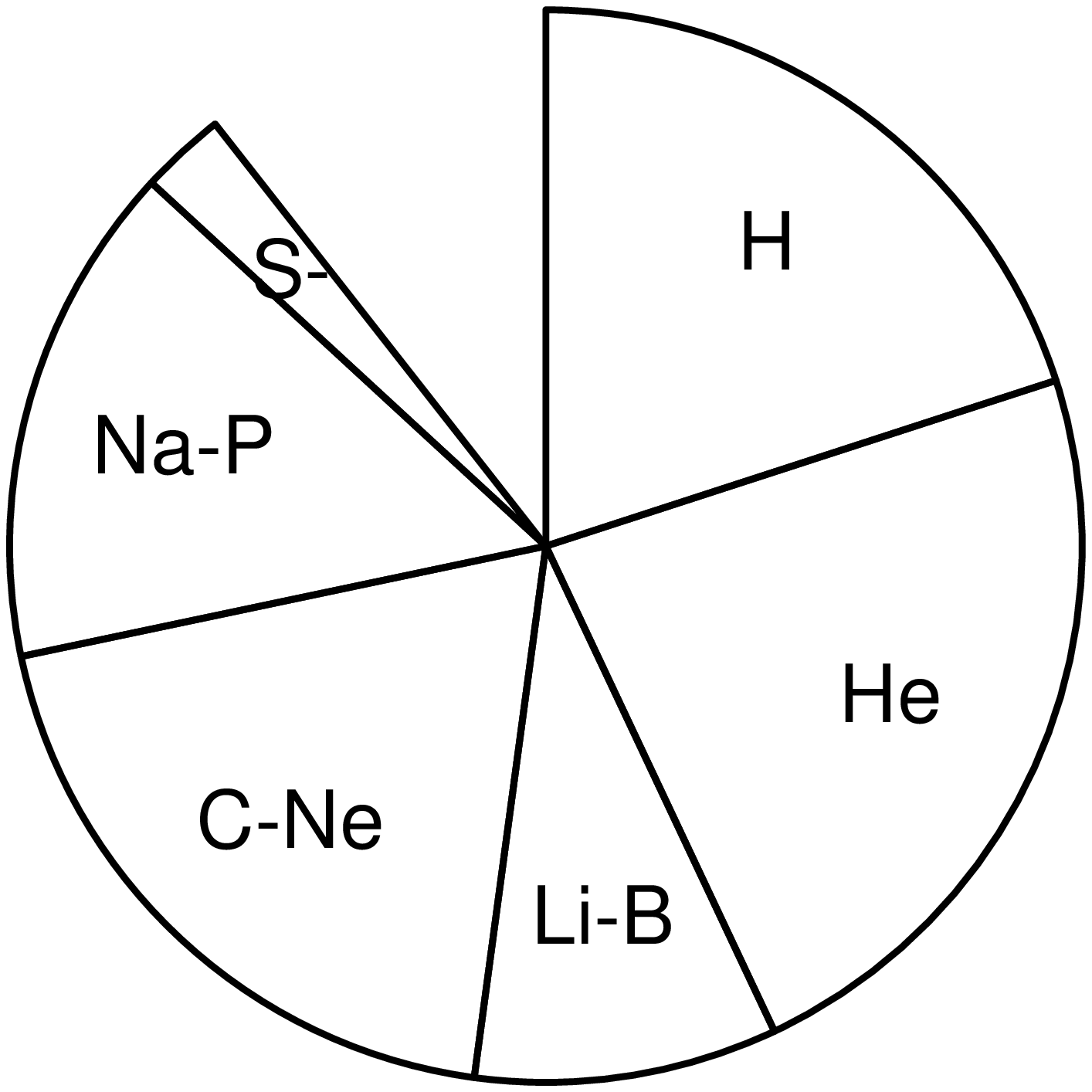}
AMD
\end{minipage}
\caption{Partition of total charge into fragments at the final state
  of the reaction $\calcium + \calcium$ at 35 MeV/nucleon: the
  experimental data of Hagel \textit{et al.}\cite{Hagel} (left) and
  the AMD result (right). }
\label{fig:chargepart}
\end{figure}

The simulations are performed in the usual way. The time evolutions
are calculated up to $t=300$ fm/$c$, when the produced fragments are
no longer strongly interacting each other. Simulations are carried out
for many ($\sim 1000$) events independently.  Figure
\ref{fig:manyeach} shows the time evolution of the density projected
on the reaction plane for several very central reaction events.
The range of the impact parameter $b_{\text{imp}}<7$ fm is
investigated, which is wide enough to compare the simulation results
with the experimental data of central reactions \cite{Hagel}. The
fragments at $t=300$ fm/$c$ are identified by the condition that two
nucleons $i$ and $j$ belong to the same fragment if
$\frac{1}{\sqrt{\nu}}|\mathbf{Z}_i-\mathbf{Z}_j|<r_{\text{frag}}$ with
$r_{\text{frag}}=5$ fm, and the decays of excited fragments are
calculated by using a statistical decay code \cite{maruyama}. To
compare the results with experimental data, the same experimental
filter and event selection as in the experiment \cite{Hagel} are
applied.  The obtained fragment charge distribution is shown in Fig.\
\ref{fig:compareEX} together with the experimental data \cite{Hagel}.
In Fig.\ \ref{fig:chargepart}, we also show how the total charge of
the system is distributed in fragments in the final state.
The experimental data show that 20\% of protons are emitted as
protons, deuterons, and tritons, 30\% of protons are contained in He
isotopes, and the rest of the protons are contained in heavier
fragments.  The features of the experimental data are reproduced by
AMD well, as we see in Figs.\ \ref{fig:compareEX} and
\ref{fig:chargepart}.  Reasonable reproduction of the fragments with
$Z=1$ and 2 is obtained.

In this paper, we have chosen the coherence time parameter $\tau_0$ to
be 5 fm/$c$ [Eq.\ (\ref{eq:tau0})]. No significant difference is seen
even if we take $\tau_0$ to be a half or twice this choice. When we
take $\tau_0$ to be much longer such as $\tau_0\sim100$ fm/$c$,
excessive production of heavy fragments is observed and consequently
the amounts of the fragments around the B-Ne region are underestimated
compared with the experimental data.  The same trend has been seen
when we use the AMD model described in Ref.\ \cite{furutaLGpt}. This
can be understood because the treatment in Ref.\ \cite{furutaLGpt}
corresponds to a relatively weak decoherence (a long coherence time)
as explained in the Appendix.

\begin{figure}
\begin{center}
\includegraphics[width=\columnwidth]{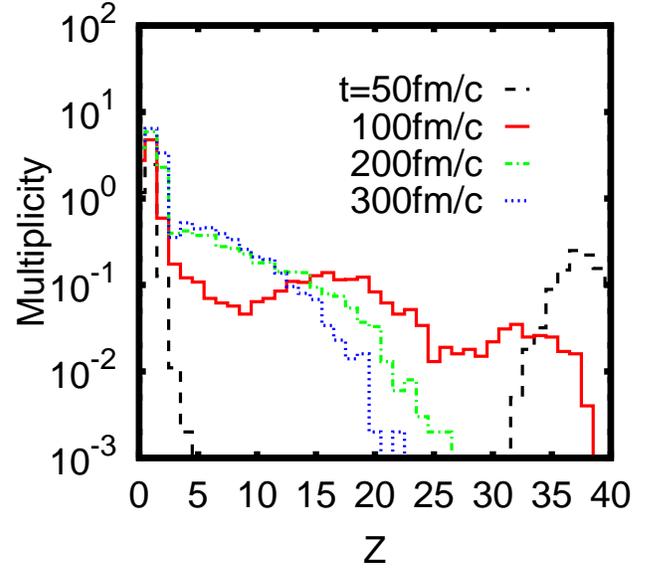}
\end{center}
\caption{(Color online) The fragment charge distribution of the very
  central ($b_{\text{imp}}=0$ fm) reaction $\calcium+\calcium$ at 35
  MeV/nucleon at four reaction times $t=$50--300 fm/$c$.}
\label{fig:timeCharge}
\end{figure}%
\begin{figure}
\begin{center}
\includegraphics[width=\columnwidth]{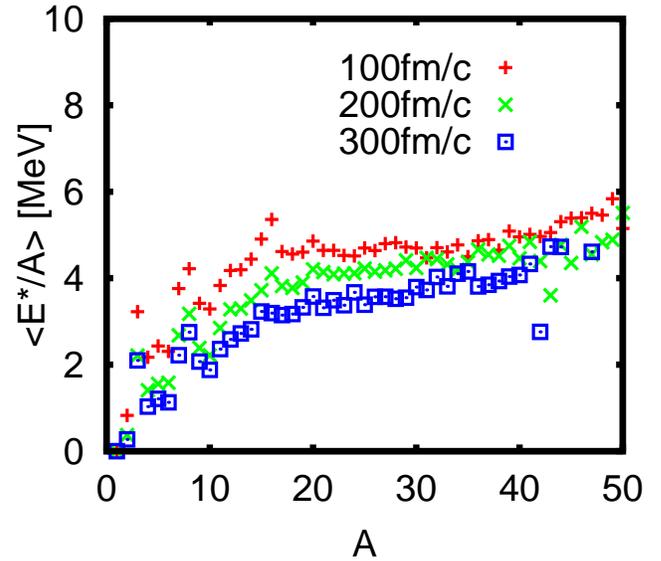}
\end{center}
\caption{(Color online) The average excitation energy of fragments as
  a function of the fragment mass number $A$ for the very central
  ($b_{\text{imp}}=0$ fm) reaction $\calcium + \calcium$ at 35
  MeV/nucleon at three reaction times $t=$100--300 fm/$c$.}
\label{fig:timeEnergy}
\end{figure}
Let us concentrate our arguments on the time evolution of several
observables for very central reaction events ($b_{\text{imp}}=0$ fm).
The fragment charge distribution and the average excitation energy as
a function of the fragment mass number are shown for several reaction
times in Figs.\ \ref{fig:timeCharge} and \ref{fig:timeEnergy},
respectively, by identifying fragments with $r_{\text{frag}}=3$ fm.
The fragments identified in this way are not necessarily related to
the fragments at the end of the reaction.  Nevertheless, these
quantities are helpful to understand the change of the reaction system
along the time evolution. (The choice of the parameter
$r_{\text{frag}}=3$ fm is taken to identify fragments even at early
stages of the reaction, but $r_{\text{frag}}=3$ fm seems to be too
small to identify the realistic fragments at time $t=300$ fm/$c$.)

Isolated nucleons and light fragments are identified even at a very
early stage of the reaction ($t=50$ fm/$c$); these are interpreted as
pre-equilibrium emissions of light particles.  The heavy fragments
$Z>20$ are negligible at the truncation time ($t=300$ fm/$c$).  The
average excitation energies per nucleon of the fragments $A\geq 15$
are as high as about 5 MeV at $t=100$ fm/$c$ and decrease to about 4
MeV at $t=300$ fm/$c$.

In many very central reaction events, the produced fragments seem to
be divided into two groups, projectile-like and target-like groups, at
the late stage of the reaction (Fig.\ \ref{fig:manyeach}). Therefore,
the two separate equilibrium systems of about half size will be more
relevant to this reaction system rather than a single large
equilibrium system, if the concept of equilibrium is relevant to this
reaction in any sense.

\section{Application to statistical calculations}
\label{EQUILIBRIUM}
We are able to study the statistical properties of many-nucleon
systems in equilibrium by using AMD as in Ref.\ \cite{furutaLGpt}. We
calculate the time evolution of the system of $A_\text{total}$
nucleons ($N_\text{total}$ neutrons and $Z_\text{total}$ protons)
confined in a spherical container of radius $r_\text{wall}$ for a long
time. We regard the $A_\text{total}$-nucleon state at each time as a
sample of an equilibrium ensemble.  The total energy $E_\text{total}$
of the system is conserved throughout the time evolution so that the
obtained ensemble is a microcanonical ensemble specified by the total
energy $E_\text{total}$, the volume $V_\text{total}=\frac{4}{3}\pi
r_\text{wall}^3$ and the number of nucleons
$A_\text{total}(Z_\text{total},N_\text{total})$. By extracting
statistical information (temperature $T$ and pressure $P$) from the
ensembles, we can construct caloric curves $T(E_\text{total},P)$.

\begin{figure}
\begin{center}
\includegraphics[width=\columnwidth]{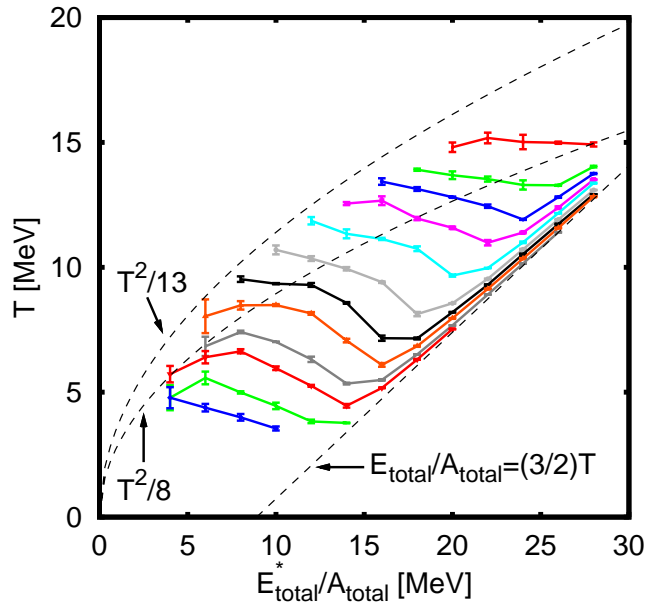}
\end{center}
\caption{(Color online) The constant-pressure caloric curves for the
  $A_\text{total}=36$ ($N_\text{total}=18$, $Z_\text{total}=18$)
  system obtained by AMD. The lines correspond to the pressure
  $P=0.005$, 0.01, 0.02, 0.03, 0.05, 0.07, 0.10, 0.15, 0.20, 0.25,
  0.30 and 0.40 MeV/fm$^3$ from the bottom upward. Statistical
  uncertainties are shown by error bars. The curves of
  $E^\ast_\text{total}/A_\text{total}=T^2/(8\,\text{MeV})$ and
  $E^\ast_\text{total}/A_\text{total}=T^2/(13\,\text{MeV})$, and the
  line of
  $E_\text{total}/A_\text{total}=(E^\ast_\text{total}+E_\text{g.s.})
  /A_\text{total}=\frac{3}{2}T$ are drawn for comparison.}
\label{fig:pconst}
\end{figure}

We utilize the same AMD model used to simulate the reaction $\calcium
+ \calcium$ at 35 MeV/nucleon in the previous section to study the
equilibrium system of $(Z_\text{total},N_\text{total})=(18,18)$, which
is the same system as studied in Ref.\ \cite{furutaLGpt}. We calculate
ensembles for various energies
$E^\ast_\text{total}/A_\text{total}=$5--8 MeV and volumes
$r_\text{wall}=$5--15 fm
($V_\text{total}/V_0=$2.5--67). $E^\ast_\text{total}$ stands for the
excitation energy relative to the ground state of the
${}^{36}\mathrm{Ar}$ nucleus ($E_\text{g.s.}=-8.9A_\text{total}$ MeV)
and $V_0=A_{\text{total}}/\rho_0$ corresponds to the volume for the
system with normal nuclear matter density $\rho_0$.  The obtained
constant-pressure caloric curves are shown in Fig.\ \ref{fig:pconst}.
Although several changes (explained in Sec.\ \ref{AMDformalism} and
the Appendix) have been introduced in this paper, the characteristic
feature of the phase transition in finite systems
\cite{grossPR,grossEP,grossPCCP}, namely negative heat capacity, can
be recognized in the constant-pressure caloric curves with $P\leq
0.3\text{MeV}$, as has been seen in Ref.\ \cite{furutaLGpt}.  The
caloric curve for the liquid phase, namely the line obtained by
connecting the leftmost points of the constant-pressure caloric
curves, shifted slightly left compared with the result of Ref.\
\cite{furutaLGpt} and, connected to that, the critical point seems to
be affected slightly. This is mostly due to the change of decoherence,
as explained in Sec. VB in Ref.\ \cite{furutaLGpt}.

The created equilibrium ensembles in this section are utilized in
Sec.\ \ref{COMPARISON}.

\section{Comparison between a dynamical simulation and statistical
  calculations}
\label{COMPARISON}
In this section, we compare two ensembles - a reaction ensemble and an
equilibrium ensemble - to study whether the concept of equilibrium is
relevant in multifragmentation reactions.
\begin{enumerate}[{(}i{)}]
\item A \textit{reaction ensemble} is obtained by collecting the
  states at a certain reaction time from many events of a dynamical
  multifragmentation reaction simulated by AMD (Sec.\
  \ref{REACTION}). The reaction ensemble is specified by the reaction
  time $t$, and we consider the ensembles obtained from the reaction
  $\calcium + \calcium$ at 35 MeV/nucleon in Sec.\ \ref{REACTION}. We
  use only very central reaction events ($b_\text{imp}=0$ fm).
\item An \textit{equilibrium ensemble} is obtained by calculating
  the time evolution of a many-nucleon system in a container for a
  long time by AMD and regarding a state at each time as a sample
  (Sec.\ \ref{EQUILIBRIUM}). The equilibrium ensemble here is a
  microcanonical ensemble specified by the total energy
  $E_\text{total}$, container volume $V_\text{total}$, and number of
  nucleons $A_\text{total}(Z_\text{total},N_\text{total})$.  We
  consider the system of $(Z_\text{total},N_\text{total})=(18,18)$
  studied in Sec.\ \ref{EQUILIBRIUM}.
\end{enumerate}
We utilize the same AMD model to calculate both situations so that we
are able to compare the reaction and equilibrium ensembles without
ambiguities.

The comparison of the reaction and equilibrium ensembles is performed
by calculating the same observables for both ensembles. In this paper,
the fragment charge distribution $Y_Z$ and the average excitation
energy as a function of the fragment mass number $\langle
E^*/A\rangle_A$ are chosen as the observables (``fragment
observables''). To make a detailed comparison with many observables,
we introduce three classes of fragment observables by choosing
different values of the fragment identification parameter
$r_{\text{frag}}$: $r_{\text{frag}(1)}=3$ fm, $r_{\text{frag}(2)}=2.5$
fm and $r_{\text{frag}(3)}=2$ fm (see Sec.\ \ref{REACTION}).  The
fragment observables with different $r_{\text{frag}}$ can be regarded
as different observables for the comparison of ensembles.

For a given reaction time $t$, we compute a quantity
\begin{align}
\delta^2&=\frac{1}{3\times13}
\biggl[ \sum_{i=1}^3\sum_{Z=4}^{16}\Bigl\{ \ln
  Y^{(i)}_{Z\;\text{react}}
  -\ln\bigl(\mathcal{N}Y^{(i)}_{Z\;\text{equil}}\bigr)\Bigr\}^
  2\biggr]\notag\\ 
&+\frac{1}{3\times14}\biggl[ 
\frac{1}{\epsilon^2}
\sum_{i=1}^3\sum_{A=2}^{15}\Bigl\{ \langle
  E^\ast/A\rangle^{(i)}_{A\;\text{react}}- \langle
  E^\ast/A\rangle^{(i)}_{A\;\text{equil}}\Bigr\}^2 \biggr]
\label{eq:deltaall}
\end{align}
and search the equilibrium ensemble that gives the minimum value of
$\delta^2$. Here $Y_Z$ and $\langle E^\ast/A\rangle_A$ are the yield
of the fragments with the charge number $Z$ and the average excitation
energy of the fragments with mass number $A$, respectively. The
subscripts ``react'' and ``equil'' indicate that the observables for
the reaction ensemble and for the equilibrium ensemble,
respectively. The superscript $(i)$ denotes that the observables are
calculated with the fragments identified by $r_{\text{frag}(i)}$.  The
factor $\epsilon$ is a dimensional constant of energy and taken as 1
MeV in this paper. The factor $\mathcal{N}$ is a normalization
constant that is optimized to give the minimum value of $\delta^2$.
The yields $Y_Z$ of the fragments with $Z=$1--3 are omitted to compute
$\delta^2$ to avoid the effect of pre-equilibrium emissions.

In Sec.\ \ref{REACTION}, we have seen that the reaction system seems
to be composed of two separate equilibrium systems if equilibrium is
relevant to this reaction. The system
$(Z_\text{total},N_\text{total})=(18,18)$ we studied in Sec.\
\ref{EQUILIBRIUM} is about half the size of the reaction system. We
therefore compare the reaction ensemble of Sec.\ \ref{REACTION} and
the equilibrium ensembles of Sec.\ \ref{EQUILIBRIUM}.  A value of
$\mathcal{N}\approx2$ is expected to compare the equilibrium system of
$(Z_\text{total},N_\text{total})=(18,18)$ with the reaction system of
$\calcium +\calcium$.  In early stages of the reaction, a heavy
fragment with $Z>20$ is identified when the projectile-like and
target-like groups overlap spatially. It is not appropriate to compare
such situations with an equilibrium ensemble with
$(Z_\text{total},N_\text{total})=(18,18)$. We therefore exclude from
the reaction ensemble the states in which a heavy fragment with $Z>20$
is identified with $r_{\text{frag}(1)}=3$ fm. We start the comparison
after the reaction time $t=80$ fm/$c$ at which we find a significant
number of adopted states.

The reaction ensembles at the time $t=80$, 100, 120, 140, 160, 180,
240 and 300 fm/$c$ are compared with the equilibrium ensembles
$E^\ast_\text{total}/A_\text{total}=$5--8 MeV and
$r_{\text{wall}}=$5--9 fm ($V_\text{total}/V_0=$2.5--14). When the
energy of the equilibrium system $E^\ast_\text{total}/A_\text{total}$
is varied, a large change is observed in $\langle E^*/A\rangle_A$.
However, when the volume of the equilibrium system $V_\text{total}$ is
varied, the change of the shape of $Y_Z$ is noticed. Therefore, by
reproducing $Y_Z$ and $\langle E^*/A\rangle_A$, we are able to find an
equilibrium ensemble (specified by $E_\text{total}$ and
$V_\text{total}$) that reproduces the reaction ensemble, if it exists.
The observables for the equilibrium ensemble depend on the size of the
system, but we have confirmed that this dependence is compensated by
the freedom of the normalization factor $\mathcal{N}$ when we change
the number of nucleons in the equilibrium system to
$A_\text{total}=20$.

\begin{figure*}[p]
\begin{center}
\includegraphics[width=0.70\textwidth]{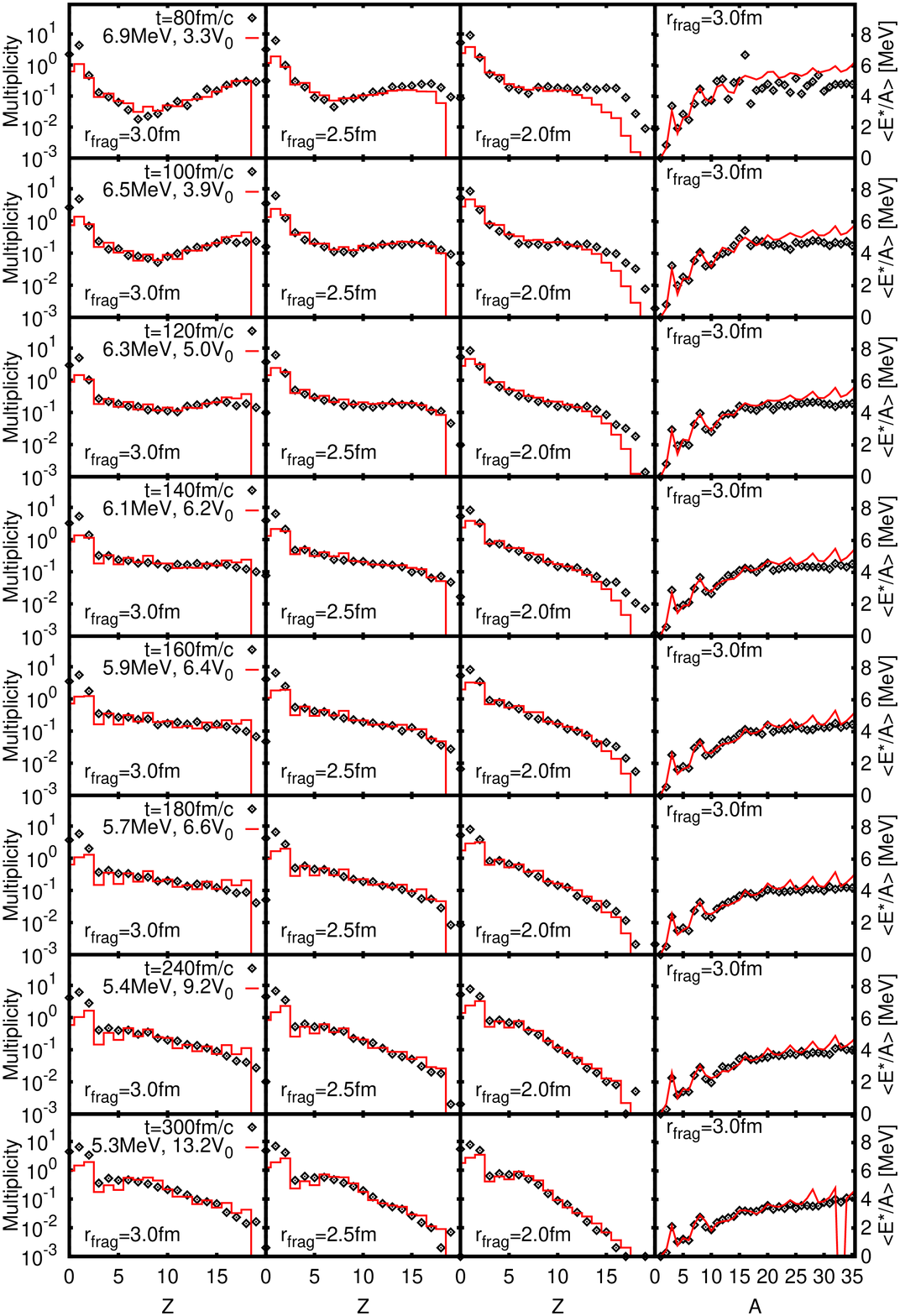}
\end{center}
\caption{(Color online) Comparison of the fragment charge distribution
  (left three columns) and the average excitation energy of fragments
  (rightmost column) of the reaction ensemble at each reaction time
  $t=$80--300 fm/$c$ obtained from the very central
  ($b_\text{imp}=0$ fm) reaction $\calcium+\calcium$ at 35 MeV/nucleon
  and those of the best-fit equilibrium ensemble of the system
  $(Z_\text{total},N_\text{total})=(18,18)$. The reaction time of the
  reaction ensemble and the energy and the volume of the equilibrium
  ensemble are shown in the leftmost column. The values of
  $r_{\text{frag}}$ for the fragment identification are indicated.}
\label{fig:fitall}
\end{figure*}

Figure \ref{fig:fitall} is the comparison of the observables between
the reaction and equilibrium ensembles. The equilibrium ensemble is
chosen to minimize the $\delta^2$ value for the reaction ensemble at
each reaction time $t=$80--300 fm/$c$. Overall features of the
fragment observables of both ensembles agree well at every reaction
time except for small details. The comparisons of $\langle
E^*/A\rangle_A$ for the fragments identified with $r_\text{frag}=2.5$
and 2 fm are not shown in Fig.\ \ref{fig:fitall}, but the agreement
between the two ensembles are as good as in the case of
$r_\text{frag}=3$ fm.
\begin{table}
\caption{
  The state variables of the equilibrium ensemble that reproduces
  the fragment observables of the reaction $\calcium+\calcium$ at
  35 MeV/nucleon at each reaction time $t=$80--300 fm/$c$.
}
\begin{center}
\begin{tabular}{ccccccc} \hline\hline
$t$ &$E^\ast_\text{total}/A_\text{total}$ & $V_\text{total}/V_0$ &$T$ &$P$ 
& $\mathcal{N}$  &$\delta^2$ \\ 
$[\text{fm}/c]$ & (MeV)   &      & (MeV) & (MeV/fm$^3$) &  &  \\
\hline
80         &6.9      &3.3    &8.1    &0.042  &1.4  &0.46 \\
100        &6.5      &3.9    &7.1    &0.029  &2.0  &0.18 \\
120        &6.3      &5.0    &6.1    &0.019  &2.1  &0.12 \\
140        &6.1      &6.2    &5.9    &0.013  &2.0  &0.13 \\
160        &5.9      &6.4    &5.6    &0.011  &2.0  &0.12 \\
180        &5.7      &6.6    &5.4    &0.010  &2.0  &0.13 \\
240        &5.4      &9.2    &4.7    &0.007  &1.8  &0.11 \\
300        &5.3      &13.2   &4.1    &0.005  &1.9  &0.15 \\\hline\hline
\end{tabular}
\end{center}
\label{tb:bestfits}
\end{table}
Table \ref{tb:bestfits} shows the energy, volume, temperature and
pressure of the best-fit equilibrium ensembles. The normalization
factor $\mathcal{N}$ and $\delta^2$ values are also listed in the
table.  The larger volume is required to reproduce the later stage of
the reaction while the energy per nucleon decreases gradually. As a
consequence, the temperature and pressure of the system decrease along
the reaction time.
\begin{figure}
\begin{center}
\includegraphics[width=\columnwidth]{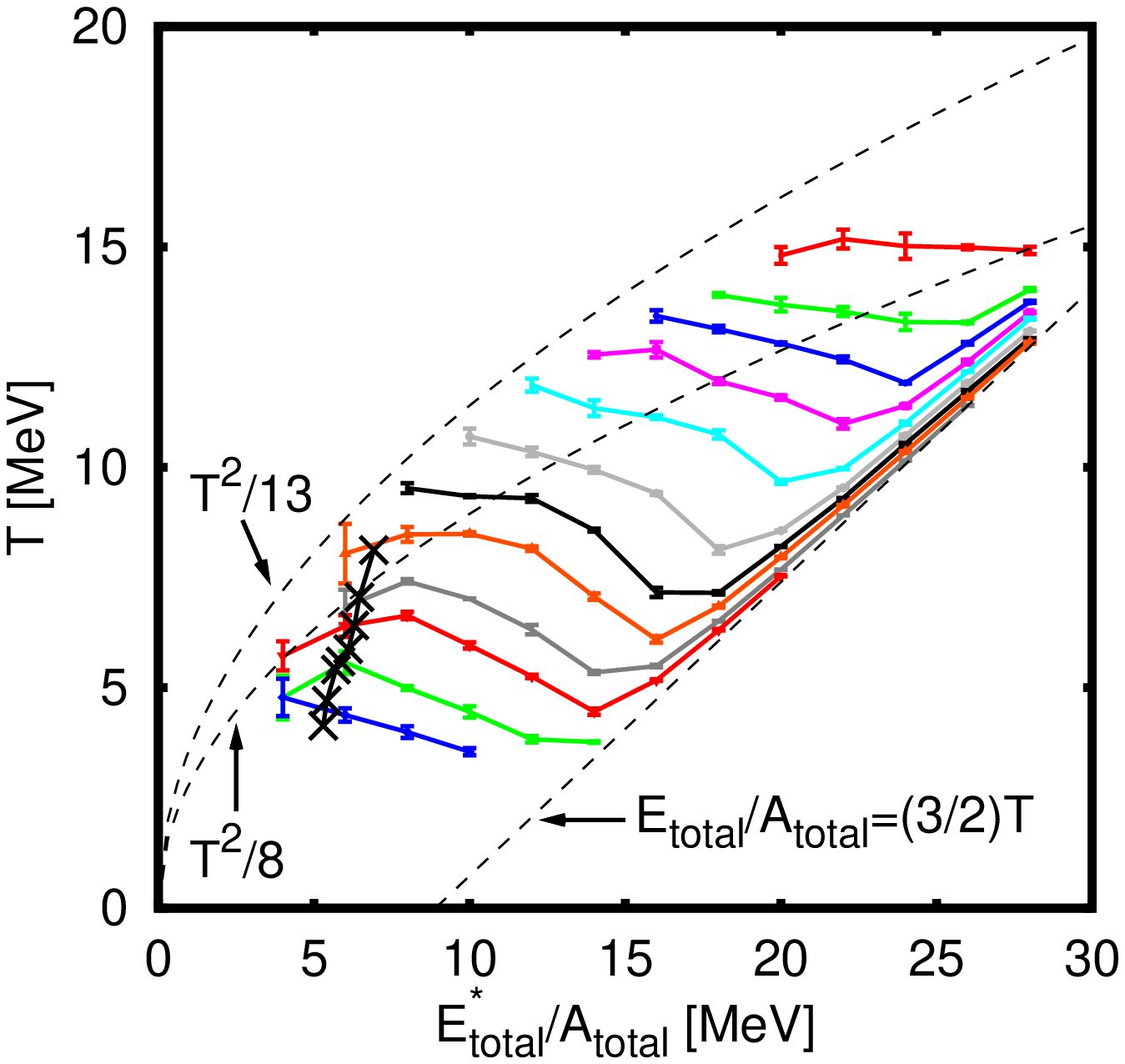}
\end{center}
\caption{(Color online) The crosses on the $E$-$T$ plane indicate the
  equilibrium ensembles that reproduce the fragment observables of the
  reaction ensembles at reaction times $t=80$, 100, 120, 140, 160,
  180, 240, and 300 from the top downward. The constant-pressure
  caloric curves of Fig.\ \ref{fig:pconst} are shown for comparison.}
\label{fig:path}
\end{figure}%
In the caloric curves of Fig.\ \ref{fig:path}, we show the reaction
path by connecting the points $(E^\ast_\text{total}/A_\text{total},T)$
of the equilibrium systems corresponding to different reaction
times. All these points seem to be located in the region of liquid-gas
coexistence that includes the region of negative heat capacity, and
therefore it seems that the fragmentation of this reaction is
connected to the nuclear liquid-gas phase transition.

These results may be interpreted as follows. The fragment observables
of the reaction system become equivalent to those of an equilibrium
system by the time $t=80$ fm/$c$ at latest. The equivalence of the
reaction and equilibrium systems is kept for a while although the
reaction system cools by breaking the fragments as well as by emitting
light fragments and nucleons. A natural question arises as to when the
equivalence between the reaction and equilibrium systems is achieved
and at what time the equivalence ends, which corresponds to the time
of freeze-out.  Unfortunately, it seems that the current choice of
observables is not suitable to discuss the beginning and the end of
the equivalence. Because the identification of fragments is impossible
at earlier stages of the reaction, it seems difficult to find
out-of-equilibrium effects after freeze-out with the resolution we
have obtained in this paper even if they exist. Even at a very late
stage such as $t=300$ fm/$c$, it seems that the fragment observables
of the reaction are still well explained by an equilibrium ensemble.

Even though the overall features match well, there are also small
discrepancies in the fragment observables between the reaction and
equilibrium ensembles.  The yields of light particles ($Z=1$ and 2) of
the equilibrium ensemble are much less than those of the reaction
ensemble, which is due to the effect of pre-equilibrium emissions of
light particles.  We also notice two systematic deviations at the
early stage of the reaction ($t\sim100$ fm/$c$), which may be due to
dynamical effects. One is the deviation in $Y_Z$ for heavy
fragments. The equilibrium ensemble overestimates these fragments when
the fragments are identified by $r_{\text{frag}(1)}=3$ fm but it
underestimates these fragments with $r_{\text{frag}(3)}=2$ fm.  The
other difference is in the value of $\langle E^*/A\rangle_A$ for heavy
fragments ($A>15$), where the equilibrium results give slightly higher
values than the reaction results, even though the values of $\langle
E^*/A\rangle_A$ for lighter fragments ($A\leq15$) for the reaction and
equilibrium ensembles match well. It will be possible to discuss
dynamical effects that exist in the reaction ensemble by further
comparison in future studies.

Let us compute other observables in both the reaction and
corresponding equilibrium ensembles. In the following calculations, we
use the fragments identified by using $r_{\text{frag}}=3$ fm.

First, we compute the kinetic observables, which should also agree in
the two ensembles if complete equilibrium is achieved in the
reaction. Unfortunately, it is not straightforward since we are
comparing the reaction system of $\calcium+\calcium$ with the
equilibrium system of $(Z_\text{total},N_\text{total})=(18,18)$ and
thus kinematics is different.  However, we are using the very central
reaction events ($b_\text{imp}=0$) and the fragments in the reaction
system seems to be categorized into two groups, projectile-like and
target-like groups, and therefore the observables related to the
transverse momentum may be little affected by the difference of
kinematics. To further reduce the influence of different kinematics,
we define the transverse direction on an event-by-event basis for the
reaction system. Choosing the $z'$-axis obtained by connecting the
center of mass of the nucleons located in the positive side of the
beam axis (the projectile-like group) and that of the nucleons located
in the negative side (the target-like group), we compute the
transverse momentum ($P_{x'},P_{y'}$) of each fragment projected on
the $x'y'$-plane perpendicular to the $z'$-axis. For the equilibrium
system, the $z'$-axis can be taken arbitrarily. We calculate the
following quantities as functions of the fragment mass number $A$;
\begin{align}
E_{\perp}(A) &=\frac{1}{2\mu(A)} \langle
P_{x'}^2+P_{y'}^2 \rangle_A\\
E_\perp^{\text{flow}}(A)& =\frac{\langle
P_\perp^\text{flow}\rangle_A^2} {2\mu(A)}
\end{align}
where the brackets $\langle\;\rangle_A$ denote the average for all the
fragments with mass number $A$ in the ensemble.  The momentum and the
position of a fragment are denoted by $P_{\sigma}$ and $R_{\sigma}$
($\sigma=x',y',z'$), respectively.  $P_\perp^\text{flow}$ is the
momentum component in the transverse radial direction
$(R_{x'},R_{y'})$;
\begin{equation}
P_\perp^\text{flow}=\frac{P_{x'}
R_{x'}+P_{y'}R_{y'}}
{\sqrt{R_{x'}^2+R_{y'}^2}}.
\end{equation}
The reduced mass $\mu(A)$ of a fragment is defined by
\begin{equation}
\frac{1}{\mu(A)}=\frac{1}{m_N}\left(
\frac{1}{A_\text{system}-A}+\frac{1}{A}\right)
\end{equation} 
where $m_N$ is the nucleon mass and $A_\text{system}$ is the
number of nucleons in the system. We take $A_\text{system}=40$ for the
reaction system since the reaction system seems to be composed of two
groups, and we take $A_\text{system}=36$ for the equilibrium
system.
\begin{figure}
\begin{center}
\includegraphics[width=0.45\textwidth]{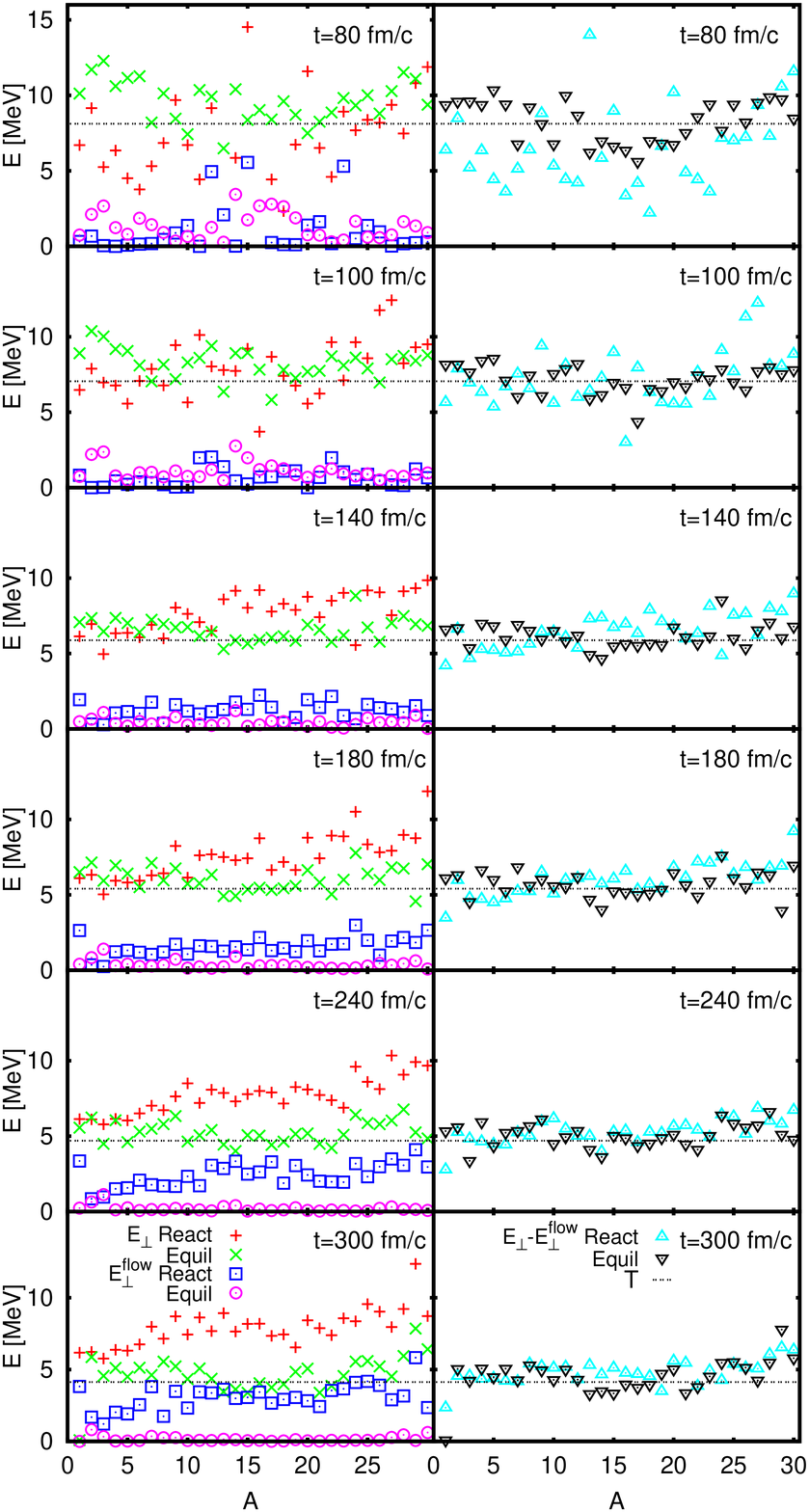}
\end{center}
\caption{(Color online) Comparison of the observables related to the
  fragment transverse momenta ($E_\perp$ and $E_\perp^\text{flow}$) of
  the reaction ensemble (``React'') at each reaction time $t=$80--300
  fm/$c$ with those of the corresponding equilibrium ensemble
  (``Equil'').  $E_\perp$ and $E_\perp^\text{flow}$ are shown in the
  left column as functions of the fragment mass number $A$, and the
  difference $E_\perp-E_\perp^\text{flow}$ is shown in the right
  column. The dashed horizontal lines indicate the temperature $T$ of
  the equilibrium ensembles.}
\label{fig:Eflowall}
\end{figure}%
The comparison between the reaction and equilibrium ensembles is shown
in the left panels of Fig.\ \ref{fig:Eflowall} for the observables
$E_\perp(A)$ and $E_\perp^{\text{flow}}(A)$ at various reaction times.
Large differences between the ensembles are found for these
observables especially at the late stage of the reaction.  For
instance, non-negligible $E_\perp^{\text{flow}}$ for the reaction
ensemble (shown by squares) is noticed at $t\gtrsim 140$ fm/$c$,
whereas $E_\perp^{\text{flow}}$ for the equilibrium (shown by circles)
is almost zero for all the times as it should be for equilibrated
systems.  (At $t=80$ fm/$c$, the statistical results are
insufficient to draw conclusions.)  However, the difference
$E_\perp(A)-E_\perp^{\text{flow}}(A)$ agrees quite well between the
reaction and equilibrium ensembles, as shown in the right panels of
Fig.\ \ref{fig:Eflowall}, at all the shown times.  Furthermore,
$E_\perp(A)-E_\perp^{\text{flow}}(A)$ has nearly no dependence on the
mass number $A$, and its value almost agrees with the value of the
temperature $T$ of the equilibrium ensemble shown by the horizontal
line at each reaction time.  This surprising agreement also suggests a
consistency of the model, since the temperature has been extracted
from an equilibrium ensemble without using the information of fragment
kinetic energies.  Thus the reaction results for the observables
related to the fragment momenta seem to be still consistent with the
equilibrium results if we subtract the flow effects from the reaction
results.

\begin{figure}
\begin{center}
\includegraphics[width=\columnwidth]{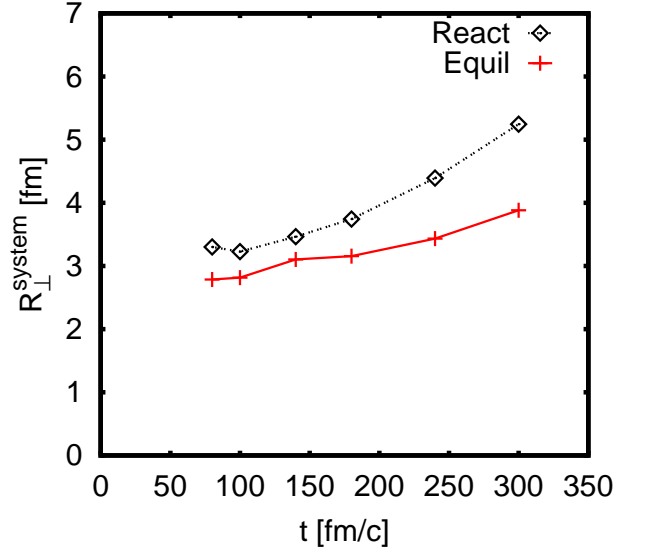}
\end{center}
\caption{(Color online) The root mean square radius
  $R_\perp^\text{system}$ of the total system in the $x'y'$-plane for
  the reaction ensemble (``React'') and the corresponding equilibrium
  ensemble (``Equil'') as functions of the reaction time.}
\label{fig:Rtot}
\end{figure}

Second, we estimate the size of the reaction system. The volume listed
in Table \ref{tb:bestfits} is that of the equilibrium system that
gives the best fit for the fragment observables of the reaction
system. However, the real volume of the reaction system may be
different from this. To estimate the size of the reaction system, we
compute the root mean square radius of the total system in the
$x'y'$-plane
\begin{equation}
R_\perp^\text{system}=\left\langle\sqrt{\frac{1}{N_{\mathcal{S}(A>5)}}
\sum_{i\in\mathcal{S}(A>5)}
(R_{ix'}^2+R_{iy'}^2)}\right\rangle
\label{eq:Rsystem}
\end{equation}
by using the nucleon positions $R_{i\sigma} (\sigma=x',y',z')$, where
$\mathcal{S}(A>5)$ denotes the nucleons that belong to the fragments
with mass number greater than 5, and $N_{\mathcal{S}(A>5)}$ is the
number of these nucleons in each event.  The nucleons that belong to
light fragments ($A\leq5$) are omitted from the calculation in Eq.\
(\ref{eq:Rsystem}) to minimize the effect of pre-equilibrium
emissions.  The results are given in Fig.\ \ref{fig:Rtot}.
The radius $R_\perp^\text{system}$ of the reaction ensemble is larger
than that of the equilibrium ensemble at all the reaction times and
the difference increases with time. For the reaction ensemble, the
system may be more extended along the beam axis owing to the memory of
reaction dynamics and then the difference of the volume between the
reaction and equilibrium systems will be more prominent. Therefore,
the difference of $R_\perp^{\text{system}}$ shown in
Fig. \ref{fig:Rtot} suggests that the real volume of the reaction
system is larger than the volume of the corresponding equilibrium
system typically by 50 \% or more.  Conversely, if the real volume is
required to agree between reaction and equilibrium ensembles, any good
fitting of $Y_Z$ will not be obtained, because of the strong volume
dependence of $Y_Z$ of the equilibrium system as can be seen from
Fig. \ref{fig:fitall}. (The dependence of $Y_Z$ on the system energy
is weak, as mentioned eqrlier.)  Thus the usual static equilibrium at
each instant is not realized.  This may be because fragments are
formed in a dynamically expanding system and the observables of
fragments recognized at a reaction time $t$ may be reflecting the
history of the state of the system before $t$ rather than the volume
at that instant $t$.

\begin{figure}
\begin{center}
\includegraphics[width=0.95\columnwidth]{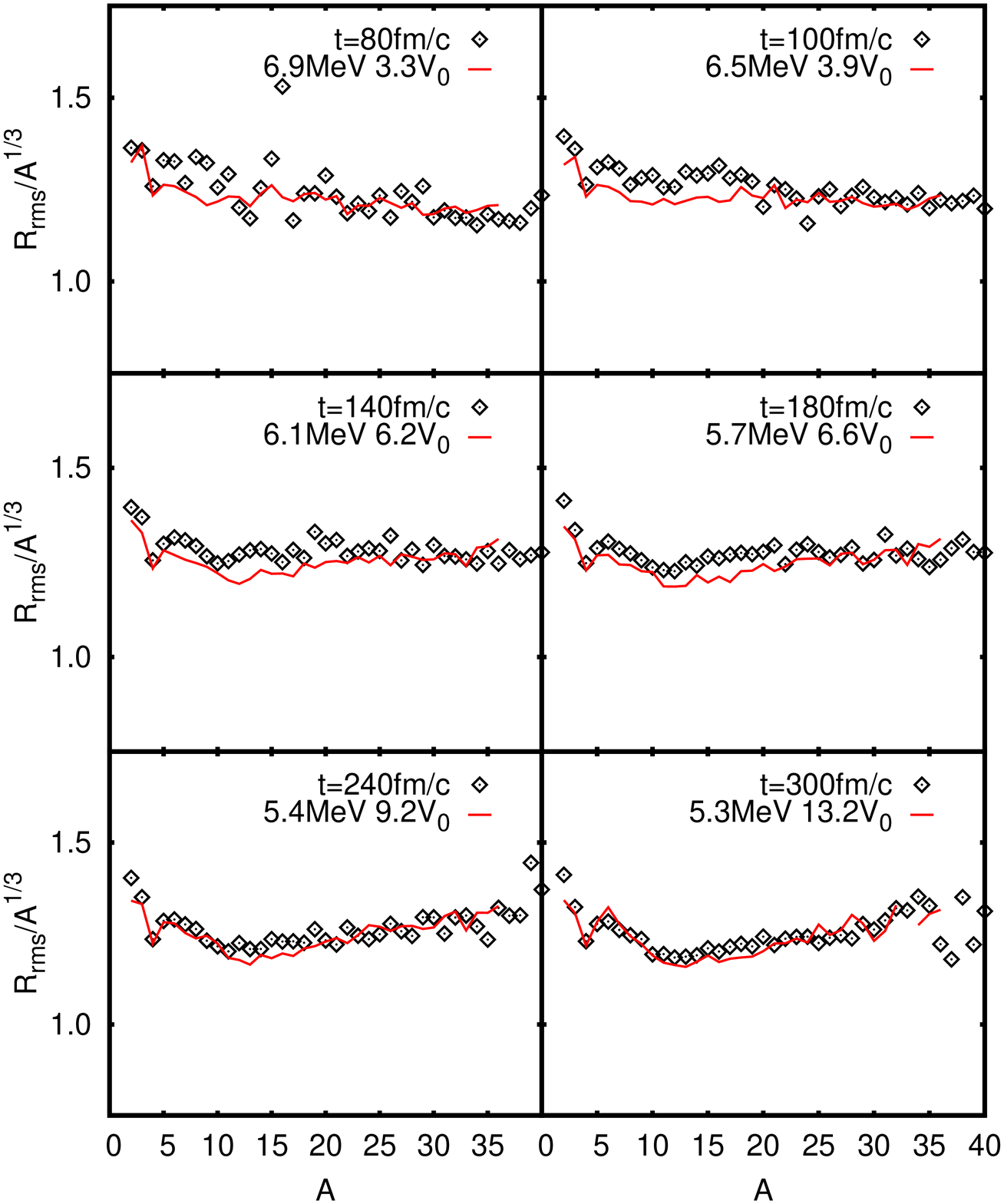}
\end{center}
\caption{(Color online) The fragment root mean square radius
  $R_\text{rms}(A)$ divided by $A^{1/3}$ plotted as a function of the
  fragment mass number $A$ for the reaction ensemble at each reaction
  time $t=$80--300 fm/$c$ and for the corresponding equilibrium
  ensemble.}
\label{fig:rmsall}
\end{figure}

Third, we calculate the root mean square radii of fragments
$R_\text{rms}(A)$ for the reaction and equilibrium ensembles to
investigate whether the properties of the created fragments in the
reaction system are the same as those in the equilibrium system (Fig.\
\ref{fig:rmsall}).
We find that $R_{\text{rms}}(A)$ of intermediate-mass fragments
($A=$6--20) for the reaction ensemble are systematically (about 5
\%) larger than those for the equilibrium ensemble.
\begin{figure}
\begin{center}
\includegraphics[width=\columnwidth]{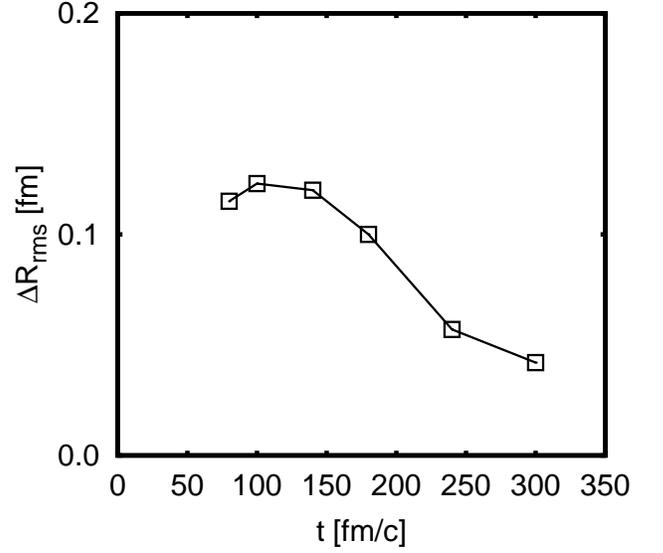}
\end{center}
\caption{The time dependence of the difference of the fragment root
  mean square radius between the reaction and corresponding
  equilibrium ensembles, averaged over the intermediate-mass fragments
  ($A=$6--20).}
\label{fig:rmsdiff}
\end{figure}%
We also calculate the average of the difference of $R_{\text{rms}}(A)$
between the ensembles over a range of intermediate mass fragments
($A=$6--20);
\begin{equation}
\Delta R_\text{rms}=\frac{1}{15}\sum_{A=6}^{20}
\left\{R^\text{react}_\text{rms}(A)-R^\text{equil}_\text{rms}(A)\right\}.
\end{equation}
We plot $\Delta R_\text{rms}$ as a function of the reaction time in
Fig.\ \ref{fig:rmsdiff}. The difference is large at the early stage of
the reaction ($t\sim100$ fm/$c$) and reduces with time at the late
stage of the reaction.  In fact, the radii of the intermediate-mass
fragments in the reaction ensemble at $t\sim100$ fm/$c$ are larger
than those in any of the equilibrium ensembles that we have
investigated.  This may be an indication of the fragment formation
mechanism in which the fragments in the reaction are made from the
expanding dilute system where surface effects are less important. It
may also be related to the finding that the symmetry energy extracted
from the multifragmentation reactions shows almost no surface effect
\cite{onoSym}.

\section{Summary}
\label{SUMMARY}
In this paper, we have investigated the relevance of the equilibrium
concept in multifragmentation by comparing reaction and equilibrium
ensembles. The reaction ensemble at each reaction time $t$ is
constructed by gathering the many-nucleon states at time $t$ in AMD
simulations of very central $\calcium+\calcium$ collisions at 35
MeV/nucleon.  The equilibrium ensemble is prepared by solving the AMD
equation of motion of a many-nucleon system
$(Z_\text{total},N_\text{total})=(18,18)$ confined in a container for
a long time. We then compare the reaction ensemble at each $t$ with
equilibrium ensembles at various conditions of volume and energy.  We
have used exactly the same AMD model in simulating both situations.
To our knowledge, this is the first work that directly compares the
multifragmentation reaction and the corresponding equilibrium system
by describing both situations with one model.

The AMD model used in this paper has been modified from that in Ref.\
\cite{furutaLGpt} to better incorporate the effect of decoherence. We
have confirmed the validity of the current version of AMD by comparing
the result of $\calcium+\calcium$ reactions at 35 MeV/nucleon with the
experimental data \cite{Hagel}. We have also confirmed that the
constant-pressure caloric curves of the equilibrium system
$(Z_\text{total},N_\text{total})=(18,18)$ constructed with the same
AMD show negative heat capacity which is the signal of the phase
transition in finite systems.

The comparison between the reaction and equilibrium ensembles has been
performed by computing the fragment charge distribution and the
average excitation energies of fragments (fragment observables) for
both ensembles.  We are able to find an equilibrium ensemble that
reproduces overall features at each reaction time $t=$80--300 fm/$c$.
For the later stage of the reaction, an equilibrium ensemble with a
larger volume and a slightly lower energy is required.  This is
consistent with the scenario that the system created by heavy-ion
collisions cools during expansion.  Unfortunately, it is difficult to
identify the beginning and the end of the equivalence between the
reaction and equilibrium systems, and it will be interesting to
further develop the study to explore these.  Experimentally, isotope
thermometers have been utilized to extract the temperature from
reactions \cite{albergo,tan}. By comparing it with the temperature
obtained by numerical simulation, it may be possible to identify the
reaction stage relevant to the experimentally obtained isotope
temperature.

The reaction ensembles have been constructed without any assumption of
thermal equilibrium.  Nevertheless, we can find an equilibrium
ensemble that is almost equivalent to the reaction ensemble as far as
the fragment observables are concerned at each reaction time after
$t=80$ fm/$c$.  This is a rather surprising result, since there are
certainly some observables that reflect the reaction dynamics.  In
fact, we have given several examples of the observables that show some
discrepancy between the reaction and corresponding equilibrium
ensembles.  The fragment transverse kinetic energies are different
from those of the equilibrium system, especially for the late stages
of the reaction. However, the difference can be explained by simple
flow effects.  If the flow effects are subtracted, the fragment
kinetic energies of the reaction system is still consistent with those
of the equilibrium system.  The size of the reaction system is larger
than that of the equilibrium system. Namely, the real volume of the
reaction system is larger than the volume assigned by fitting the
fragment observables.  The difference becomes larger at the later
stages of the reaction. The usual static equilibrium at each instant
is not realized since any equilibrium ensemble with the same volume as
that of the reaction system cannot reproduce the fragment observables.
The fragment radii in the reaction system are larger than those in the
equilibrium system. The difference is large at the early stage of the
reaction ($\sim100$ fm/$c$) and decreases with time. This may be an
indication of a fragment formation mechanism in which the fragments
are made from an expanding dilute system in the reaction.

Only a small difference between the reaction and equilibrium ensembles
is seen in the fragment observables studied in this paper. However,
dynamical effects may become essential even for the fragment
observables when the incident energy is increased or the impact
parameter is varied.  It has been suggested that neck formation play
an important role in semiperipheral collisions \cite{ditoroWCI}, but
in this paper we ignored this effect.  It is an interesting question
whether the equivalence between the multifragmentation reaction and
the equilibrium system still holds under such circumstances.  It is
also interesting to compare observables such as the momentum
distribution of fragments and the system size of multifragmentation
reactions with those of the corresponding equilibrium systems in the
explicit presence of expansion and flow effects
\cite{gulminelliFlow,isonFlow}.

In this paper, we studied only one particular reaction, namely very
central $\calcium + \calcium$ collisions at 35 MeV/nucleon. The
reaction mechanism changes from one reaction to another. It is
therefore interesting to apply the same approach to other reactions,
such as a reaction of heavier nuclei where creation of a single
thermal source is expected \cite{radutaEQ,radutaEQ2}, and a reaction
of nuclei with different isospin compositions where the occurrence of
isospin diffusion has been claimed\cite{tsangID,chen}.  It is
important to explore the effects of various reaction parameters such
as the reaction system, the incident energy and the impact parameter
on the achievement of equilibrium. If the concept of equilibrium is
relevant, it is interesting to explore how these parameters influence
the parameters to specify the equilibrium system.  This study will
offer guidelines for combining experimental data of various heavy-ion
collisions to construct, for example, equation of states and
constant-pressure caloric curves.

\begin{acknowledgments}
The major part of this work has been done at Tohoku University as the
Ph.~D. study of T.F. T.F. acknowledges partial support from the ANR
project NExEN (ANR-07-BLAN-0256-02).  This work is partly supported by
the High Energy Accelerator Research Organization (KEK) as a
supercomputer project.
\end{acknowledgments}

\appendix
\section*{APPENDIX: Improved implementation of decoherence}
The reaction $\calcium+\calcium$ at 35 MeV/nucleon has already been
studied by AMD \cite{onoAMD-V,wada1998} and it has been shown that
several aspects of the experimental data \cite{Hagel} are nicely
reproduced. The AMD model used in these studies adopts the
instantaneous decoherence of the single-particle motion
\cite{onoAMD-V}. In contrast, in Ref.\ \cite{furutaLGpt} and in this
paper, we utilize the AMD model in which the coherence of the
single-particle motions are kept for a finite duration. When we
directly applied the AMD formalism given in Ref.\ \cite{furutaLGpt} to
the reaction $\calcium+\calcium$ at 35 MeV/nucleon, excessive
productions of heavy fragments are obtained and, connected to that,
amounts of lighter fragments around the B-Ne region are underestimated
compared with the experimental data. This is because the coherence
time chosen by the formalism in Ref.\ \cite{furutaLGpt} is too long
and the effect of decoherence is hindered for some cases, and thus it
fails to give enough quantum fluctuations to break the heavy
fragments.  A modification is necessary to better incorporate the
effect of decoherence and reproduce the experimental data. This is
rather technical but the summary is given in this Appendix.

In the AMD formalism, special care is taken for the nucleons that are
almost isolated. For instance, the zero-point kinetic energies of
these nucleons are subtracted since the wave functions of such
nucleons should have sharp momentum distributions rather than Gaussian
ones corresponding to the wave packet in Eq.\
(\ref{eq:Gaussian}). This change of interpretation is necessary for
the consistency of Q-values of nucleon emissions and fragmentation
\cite{onoPTP,onoPRL,onoReview} and is very important for the
definition of temperature \cite{furutaLGpt}.  In Ref.\
\cite{furutaLGpt}, we judge the ``degree of isolation'' of the nucleon
$k$ by introducing
\begin{equation}
\mathcal{I}_k=[1-w(q_k)]\mathcal{I}^{(0)}_k+w(q_k),
\label{eq:Iold}
\end{equation}
where $q_k$ counts the neighboring nucleons of the nucleon $k$
including itself, $w(q)$ is a continuous function from one when the
number of neighboring nucleons $q_k$ is small ($q\lesssim2.5$) to
zero, and $\mathcal{I}^{(0)}_k$ corresponds to the inverse number of
the neighboring nucleons.  Detailed definitions of these functions are
given in Appendix A in Ref.\ \cite{furutaLGpt}.

In the AMD formalism, the phase-space distribution $g(x;X,S)$ is
considered to compute the time evolution of the mean-field
propagation. The distribution for each nucleon $k$ is parametrized by
\begin{multline}
g(x;X_k,S_k)=\frac{1}{8\sqrt{\det S}}\\
\times\exp{\Bigl[ -\frac{1}{2}\displaystyle\sum_{a,b=1}^6
      S_{kab}^{-1}(x_a-X_a)(x_b-X_b)\Bigr]},
\label{eq:deformedWP}
\end{multline}
where $x$ gives the six-dimensional phase-space coordinates
\begin{equation}
x=\{x_a\}_{a=1,\dots,6}=\left\{\sqrt{\nu}\mathbf{r},
\;\frac{\mathbf{p}}{2\hbar\sqrt{\nu}}\right\},
\end{equation}
and $S_k$ and $X_k$ specify the shape and the centroids of the
distribution, respectively. $X_k$ is identified with the physical
coordinate $\mathbf{W}_k$ \cite{onoPRL,onoPTP,onoReview}:
\begin{equation}
X_k=\{X_{ka}\}_{a=1,\dots,6}=
\left\{\text{Re}\mathbf{W}_k,\;\text{Im}\mathbf{W}_k\right\}.
\end{equation}
In Ref.\ \cite{furutaLGpt}, one condition was imposed on
$g(x;X_k,S_k)$ for each nucleon $k$ by using the degree of isolation
$\mathcal{I}_k$.  The condition was
\begin{equation}
\mathop{\mathrm{Tr}_p} S_k\leq\frac{3}{4} (1-\mathcal{I}_k),
\label{eq:oldCONWIDTH}
\end{equation}
where $\mathop{\mathrm{Tr}_p}S_k=S_{k44}+S_{k55}+S_{k66}$ denotes the
momentum spreading of the distribution $g(x;X_k,S_k)$. Namely, if the
left-hand side of Eq.\ (\ref{eq:oldCONWIDTH}) is getting larger than
the right-hand side, $S_k$ was reduced to satisfy the equality of Eq.\
(\ref{eq:oldCONWIDTH}) and the reduced part was converted into a
stochastic Gaussian fluctuation to the centroid $X_k$ (the details are
explained in Sec. III in Ref.\ \cite{furutaLGpt}). The purpose of this
condition is to ensure full consistency of the energy conservation and
to allow precise evaluation of the temperature. When a recovery of the
phase-space distribution $g(x;X,S)$ for nucleon $k$ took place as a
result of decoherence, we replaced the shape of the distribution $S_k$
with
\begin{equation}
S_{kab}=
\begin{cases}
\frac{1}{4} & (a=b=1,2,3)\\
\frac{1}{4}(1-\mathcal{I}_k) & (a=b=4,5,6)\\
0 & (a\neq b)
\end{cases},
\label{eq:recover}
\end{equation}
where the momentum widths were chosen to be
$\frac{1}{4}(1-\mathcal{I}_k)$ rather than the standard Gaussian width
1/4 to satisfy Eq.\ (\ref{eq:oldCONWIDTH}).

The condition is arbitrary as long as $\mathop{\mathrm{Tr}_p}
S_k\simeq\frac{3}{4} (1-\mathcal{I}_k)$ is satisfied for the nucleon
$k$ that is utilized to measure the temperature. Unfortunately, it
turns out that the condition (\ref{eq:oldCONWIDTH}) utilized in Ref.\
\cite{furutaLGpt} tends to hinder the effect of decoherence
unphysically at the surface of fragments. This is because
$\mathcal{I}_k$ increases close to unity when the nucleon $k$ is
located near the surface of the fragment to which the nucleon $k$
belongs.  The increase of $\mathcal{I}_k$ results in keeping
$\mathop{\mathrm{Tr}_p} S_k$ small, even though the recovery of the
phase-space distribution defined by Eq.\ (\ref{eq:recover}) frequently
occurs. There is no physical reason why the effect of decoherence is
suppressed at the surface of fragments and it is more natural that the
effect of decoherence for the nucleon $k$ is as large as those for the
other nucleons belonging to the same fragment even though the nucleon
$k$ is located near the surface.  We thus introduce a new function
\begin{equation}
\mathcal{I}^\ast_k=w(q_k)[1-w(q_k)]\mathcal{I}^{(0)}_k+w(q_k)
\label{eq:Inew}
\end{equation}
and replace $\mathcal{I}_k$ in Eq.\ (\ref{eq:oldCONWIDTH}) and Eq.\
(\ref{eq:recover}) with this newly defined function
$\mathcal{I}^\ast_k$, while we keep $\mathcal{I}_k$, which appears in
the equation of motion as it is (see Sec. III in Ref.\
\cite{furutaLGpt}). The difference between $\mathcal{I}^\ast_k$ and
$\mathcal{I}_k$ is only that the first term of Eq.\ (\ref{eq:Iold}) is
multiplied by $w(q_k)$ so that $\mathcal{I}^\ast_k\sim0$ when the
nucleon $k$ is located inside of a fragment, whereas
$\mathcal{I}^\ast_k\sim\mathcal{I}_k$ when the nucleon $k$ has only a
few neighboring nucleons.  In addition to this modification, we change
the criteria to choose the nucleons that are used to measure
temperature of the system.  It has been shown that, to calculate the
temperature of the system correctly, it is necessary to choose the
subsystem consisting of the nucleons with negligible quantum effects
among them based on only the nucleon spatial coordinates without using
momentum variables (see Appendix B in Ref.\ \cite{furutaLGpt}).  For
this purpose, there was a condition that the nucleons that are used to
measure the temperature are chosen not to have more than one other
nucleon within a distance of 3 fm in Ref.\ \cite{furutaLGpt}. We
replace this condition with $\{k; w(q_k)\geq0.9\}$, which has similar
meaning to the aforementioned condition and guarantees that the
difference between $\mathcal{I}^\ast_k$ and $\mathcal{I}_k$ for these
nucleons is $1\%$ at most. This replacement is justified by the study
that the measured temperatures are independent of the choice of
nucleons utilized to measure the temperature as long as necessary
conditions are satisfied (see Sec.\ VC in Ref.\ \cite{furutaLGpt}).

\end{document}